\documentclass{nature}
\usepackage{eqnarray}
\usepackage[english]{babel} 
\usepackage[]{layout}
\usepackage{amsmath,amsfonts,amssymb}
\usepackage{graphicx}
\usepackage{float}
\usepackage{color}
\usepackage{braket}
\usepackage{array,multirow,makecell}
\makeatletter
\let\saved@includegraphics\includegraphics
\AtBeginDocument{\let\includegraphics\saved@includegraphics}
\renewenvironment*{figure}{\@float{figure}}{\end@float}
\makeatother

\title{Direct measurement of the quantum geometric tensor in a two-dimensional continuous medium}

\author{A.~Gianfrate$^1$, O.~Bleu$^2$\footnote{These authors contributed equally: A.~Gianfrate, O.~Bleu}, L.~Dominici$^1$, V.~Ardizzone$^1$, M.~De Giorgi$^1$, D.~Ballarini$^1$, K.~West$^3$, L.~N.~Pfeiffer$^3$, D.~D.~Solnyshkov$^2$, D.~Sanvitto$^1$, G.~Malpuech$^2$}
\begin{document}
\maketitle

\begin{affiliations}
\item CNR NANOTEC, Istituto di Nanotecnologia, via Monteroni, 73100 Lecce, Italy.
\item Institut Pascal, PHOTON-N2, Universit\'e Clermont Auvergne, CNRS, SIGMA Clermont, Institut Pascal, F-63000 Clermont-Ferrand, France.
\item PRISM, Princeton Institute for the Science and Technology of Materials, Princeton University,
Princeton, New Jersey 08540, USA
\end{affiliations}

\begin{abstract}
Topological Physics relies on the specific structure of the eigenstates of Hamiltonians. Their geometry is encoded in the quantum geometric tensor \cite{berry1989quantum} containing both the celebrated Berry curvature \cite{berry1984quantal}, crucial for topological matter \cite{Hasan2010}, and the quantum metric \cite{provost1980riemannian}.
The latter is at the heart of a growing number of physical phenomena such as superfluidity in flat bands \cite{peotta2015superfluidity}, orbital magnetic susceptibility \cite{Gao2014,Piechon2016}, exciton Lamb shift \cite{PhysRevLett.115.166802}, and non-adiabatic corrections to the anomalous Hall effect \cite{Gao2014,Bleu2018effective}.
Here, we report the first direct measurement of both Berry curvature and quantum metric in a two-dimensional continuous medium. The studied platform is a planar microcavity of extremely high finesse, in the strong coupling regime \cite{Microcavities}. It hosts mixed exciton-photon modes (exciton-polaritons) subject to photonic spin-orbit-coupling \cite{Kavokin2005} which makes emerge Dirac cones \cite{Tercas2014} and exciton Zeeman splitting breaking time-reversal symmetry. The monopolar and half-skyrmion pseudospin textures are measured by polarisation-resolved photoluminescence. The associated quantum geometry of the bands is straightforwardly extracted from these measurements. Our results unveil the intrinsic chirality of photonic modes which is at the basis of topological photonics \cite{Haldane2008,lu2014topological}. This technique can be extended to measure Bloch band geometries in artificial lattices \cite{Bleu2018}. The use of exciton-polaritons (interacting photons) opens wide perspectives for future studies of quantum fluid physics in topological systems. 
\end{abstract}

\maketitle

One of the most striking manifestations of topological effects in Physics is the conductance quantization in the two-dimensional quantum Hall effect (QHE). 
This perfect quantization relies on the existence of a topological invariant characterising the global band properties: the Chern number. Non-zero Chern numbers are also associated with the chiral conducting edge states in topological insulators and superconductors \cite{Hasan2010}. Beyond electronic systems, topological band concepts have been extended to a wide variety of wave systems covering photonics \cite{lu2014topological}, acoustics \cite{Chernacoustics2015}, cold atoms  \cite{cooper2018topological}, and even geophysics \cite{delplace2017topological}.

Topological effects are not encoded in the energy spectrum of the system but rely on the non-trivial geometry of the corresponding eigenstates.
It is the gauge invariant quantum geometric tensor that contains the structural information about the eigenstates of a parametrised Hermitian Hamiltonian. The geometric tensor is comprised of two fundamental parts: its symmetric real part that defines the quantum metric allowing to characterise distances between neighbouring states \cite{provost1980riemannian} in the parameter space, whereas its antisymmetric imaginary part determines the Berry curvature \cite{berry1984quantal}. On the one hand, the Berry curvature of the momentum space is crucial in modern Physics. Locally, it is responsible for the anomalous Hall transport \cite{MacDonald2010} in the intrinsic spin Hall effect \cite{Sinova2015} and of the valley Hall effect \cite{mak2014valley}. Its integral over a 2D closed manifold gives the Chern number itself. On the other hand, the quantum metric, closely related to the concept of fidelity used in quantum information theory, is also associated with important physical phenomena. It can be used to probe quantum phase transitions when defined in an arbitrary parameter space \cite{Zanardi2007}. The metric of the \emph{momentum} space  affects the electronic orbital magnetic susceptibility \cite{Gao2014,Piechon2016} in crystals and the exciton Lamb shift in Transitional Metal Dichalcogenides \cite{PhysRevLett.115.166802}. It has also been used to characterise superfluidity in flat bands \cite{peotta2015superfluidity} and to introduce corrections to the semiclassical equation for the anomalous Hall effect \cite{Gao2014,Bleu2018effective}. 

The extension of the use of topological concepts from solid state physics to many other classical or quantum systems in the last decade has opened new possibilities for the measurement of local geometrical properties of bands, and not only the global ones (like the conductivity in QHE). Several protocols have been proposed to measure the Berry curvature in such systems  \cite{Hauke2014,Montambaux2015}.  
Recently, experimental reconstruction of the Berry curvature via indirect dynamical measurements has been reported in artificial lattices based on quenching in a cold atom setup \cite{flaschner2016experimental} and on anomalous transport in an optical fiber based photonic mesh lattice \cite{wimmer2017experimental}. In this work, we present a direct measurement of the quantum geometric tensor components (Berry curvature and quantum metric) in an homogeneous 2D system (without periodic lattice). The experimental platform consists of a radiative photonic system where the eigenstates are exactly determined by a basic optical technique: polarisation-resolved photoluminescence (PL). The sample under study is a high quality planar microcavity ($Q>10^5$) with embedded quantum wells. The eigenmodes are  two dimensional strongly coupled exciton-photon states (polaritons) \cite{Microcavities}. Each branch is doubly polarisation degenerate and forms a pseudo-spinor. The polarisation degeneracy is lifted by the photonic splitting between the TE and TM eigenmodes \cite{Kavokin2005} (Transverse-Electric and Transverse-Magnetic) and, under magnetic field by the exciton Zeeman splitting. The polarised polariton branches therefore get entirely split, exhibiting both a non-zero Berry curvature and quantum metric that we can optically assess and characterize as discussed in the following.\\

\begin{figure}
 \includegraphics[width=1.0\linewidth]{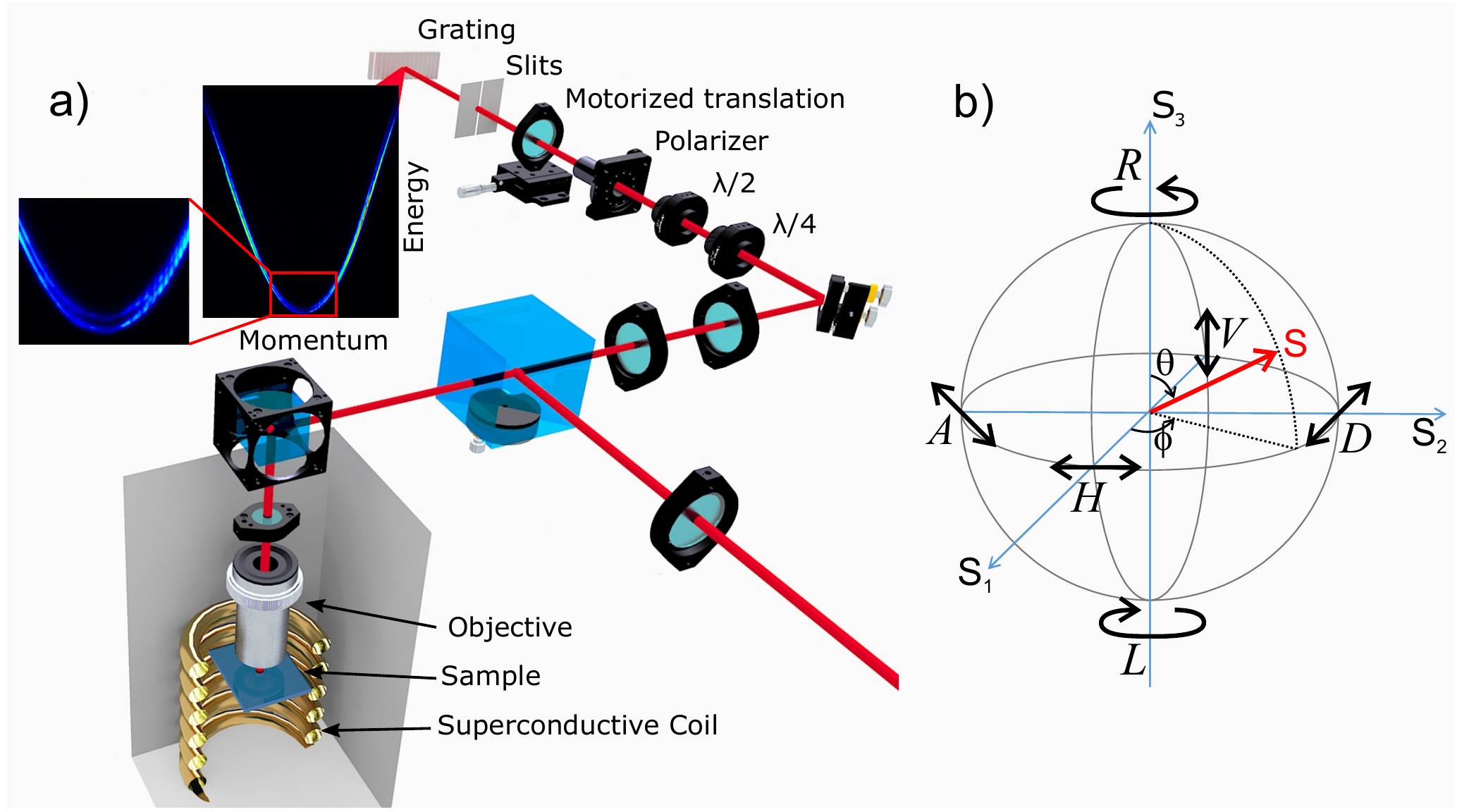}
 \caption{\label{fig1} \textbf{Experimental setup and the Poincar\'e sphere}
\small a) Scheme of the polarisation tomography experiment. The incoming pump laser (bottom right) is focused onto the sample held in the cryogenic superconductive magnet (bottom left). The sample emission is recollected, polarization filtered and the momentum space optically rebuilt at the entrance slits of a grating (top) which can resolve in energy the polariton wavefunction (top left), the Zeeman splitting due to applied magnetic field is highlighted in the inset. b) Pseudospin on the Poincar\'e sphere. The pseudospin coordinates $S_1$, $S_2$,  $S_3$ are determined by measuring the polarisation degree of emission of one state on the Horizontal/Vertical (HV), Diagonal/Anti-diagonal (DA), and circular Right/Left (RL) polarisation basis [Eq.~\eqref{pseudo}]. The $k$-dependence of the pseudospin orientation (defined by the angles $\theta$ and $\phi$) allows to compute the Berry curvature and the quantum metric  using Eqs.~\eqref{xtract}.
  }
\end{figure}

Before presenting the measured data, let us remind the main features of the specific system under study based on its description in terms of an effective two-band Hamiltonian \cite{Bleu2018} which describes the lower polariton branch with polarisation degree of freedom \cite{Microcavities}. Such description neglects the decaying nature of the radiative photonic modes (non-hermiticity), which is however only a weak perturbation in high quality samples as in our case. This effective Hamiltonian written on the basis of the circularly polarised eigenstates for small wavevectors reads:
\begin{eqnarray}
H_{\mathbf{k}}=\begin{pmatrix}
\frac{\hbar^2k^2}{2m^*}+\Delta_z &\alpha e^{-i\varphi_0}+\beta k^2 e^{2i\varphi} \\
\alpha e^{i\varphi_0}+\beta k^2 e^{-2i\varphi}& \frac{\hbar^2k^2}{2m^*}-\Delta_z
\label{Hameff}
\end{pmatrix} 
\end{eqnarray}
where $m^*=m_lm_t/(m_l+m_t)$, with $m_l$ and $m_t$ corresponding to the longitudinal and transverse effective masses and $k=|\mathbf{k}|=\sqrt{k_H^2+k_V^2}$ is the in-plane wavevector ($k_H=k\cos{\varphi}$, $k_V=k\sin{\varphi}$, and $\varphi$ is the in-plane propagation angle). 
As any $2\times 2$ Hermitian Hamiltonian, it is a linear combination of Pauli matrices, and can be interpreted as describing the interaction between an effective magnetic field and a pseudospin:
 \begin{equation}
H_{\mathbf{k}}=\frac{\hbar^2k^2}{2m^*}\mathbb{I}+ \boldsymbol{\Omega}(\mathbf{k}).\boldsymbol{\sigma}
\end{equation} 
In the present case, $\mathbf{S}=\braket{\mathbf{\sigma}}$ is nothing but the polarisation pseudospin of light, and the effective field is: 
\begin{equation}
\boldsymbol{\Omega}(\mathbf{k})=\begin{pmatrix}\alpha \cos\varphi_0+ \beta k^2\cos 2\varphi \\ \alpha \sin\varphi_0- \beta k^2\sin 2\varphi \\ \Delta_z\end{pmatrix}
\end{equation} 
Here, $\alpha$, $\beta$, and $\Delta_z$ define the strength of the effective fields corresponding to a constant splitting between X and Y polarizations ($\varphi_0$ corresponds to the in-plane angle of the constant XY field with respect to a reference H-axis), the TE-TM splitting, and the Zeeman splitting, respectively. The beautiful feature of the effective magnetic field representation is that the eigenstates have their pseudospin parallel and anti-parallel with the effective magnetic field vector. The components of the quantum geometric tensor are linked with the variation of the orientation of pseudospin vector in $k$-space. All these components cancel if $\beta=0$, namely if the TE-TM spin-orbit coupling vanishes. The Berry curvature and quantum metric  read: 
\begin{eqnarray}
g_{ij} =\frac{1}{4}(\partial_{k_i}\theta\partial_{k_j}\theta+\sin^2\theta \partial_{k_i}\phi\partial_{k_j}\phi) \nonumber\\
B_z =\frac{1}{2}\sin\theta(\partial_{k_H}\theta\partial_{k_V}\phi-\partial_{k_V}\theta\partial_{k_H}\phi)
\label{xtract}
\end{eqnarray}
with $g_{ij}$ the metric components and $B_z$ the Berry curvature.  $\theta(\mathbf{k})$ and $\phi(\mathbf{k})$ are polar and azimuthal angles which parametrise the eigenstate $\psi=\left( \cos\frac{\theta}{2} e^{-i \phi},\sin\frac{\theta}{2}\right)^T$ and the pseudospin vector position on the Poincar\'{e} sphere (Fig.~\ref{fig1}(b)). 
These quantities can be computed analytically\cite{Bleu2018} by using the effective Hamiltonian \eqref{Hameff} (the corresponding expressions are reminded in the Supplementary Note 1).

The sample studied is a high-quality polariton microcavity with a lifetime exceeding 100 ps (see Methods). The experimental setup is shown in Fig.~\ref{fig1}(a). The measurements are executed in a reflection configuration with the sample kept at cryogenic temperature and under the appliance of an external magnetic field. An off-resonant continuous wave laser is used to excite the sample inside the cryostat. The emission from the polariton modes is collected externally and its polarization tomography performed by means of half- and quarter-waveplates plus a polarizer, and the momentum space optically imaged onto the entrance slits of a monochromator. The PL is measured versus the 2D wave vector and energy for all 6 polarisation axis of the Poincar\'{e} sphere shown in Fig.~\ref{fig1}(b), which corresponds to circular left (L), circular right (R), horizontal (H), vertical (V), diagonal (D), anti-diagonal (A). For each wave vector, the two polarisation mode energies are found by fitting the total emitted intensity by a double Gaussian, which allows a good extraction even when the mode splitting is comparable with the linewidth (see Methods). Once the energy branches $E_{\pm}(\mathbf{k})$ are found, their associated pseudospin is determined from the polarisation intensities as:
\begin{equation}
S_1(\mathbf{k}) =\frac{I_H-I_V}{I_H+I_V},\qquad  
S_2 (\mathbf{k})=\frac{I_D-I_A}{I_D+I_A},\qquad
S_3(\mathbf{k}) =\frac{I_R-I_L}{I_R+I_L}
\label{pseudo}
\end{equation}
This allows to access the complete pseudospin distribution in k-space for each branch. Note that ideally, the polarisation map of one branch should be complementary with respect to the other branch, which is relatively well verified as we show below. Once the pseudospin pattern is known, the quantum metric and the Berry curvature can be computed using Eq. \eqref{xtract} with $\theta =\arccos{S_3}$ and $ \phi=\arctan S_2/S_1$.

\begin{figure}
 \includegraphics[width=1.0\linewidth]{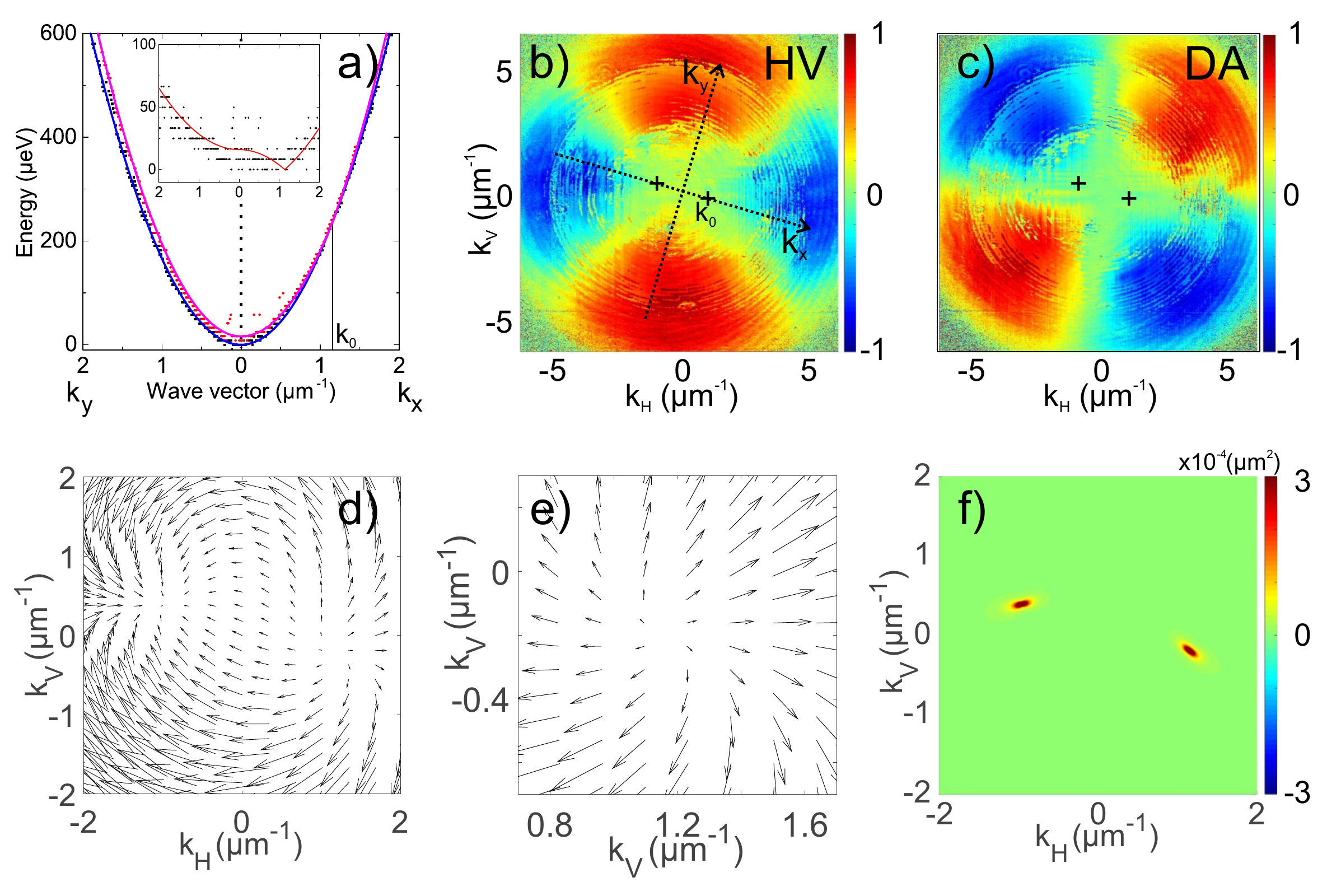}
 \caption{\label{fig2} \textbf{Emergence of pseudospin monopoles}
\small  All data presented in this figure are extracted from PL measurements taken at 0~T.
  a) The two eigenmodes energies along the two orthogonal momentum directions $k_x$ and $k_y$, shown on the two sides of the panel. Inset: eigenmodes energy splitting ($E_+-E_-$). Points - experimental data. Solid lines - parabolic fit. (b,c) HV, and DA polarisation degrees maps of the lower energy mode in the full 2D momentum space. The crosses mark the degeneracy points, where the eigenmodes are crossing (the line passing by the two crosses represent the $k_x$ direction of panel a). (d,e) The corresponding in-plane pseudospin $(S_1,S_2)$ texture distribution in $k$-space, shown on a wide scale (d), and zoomed on one of the two points (e), demonstrating a monopole pseudospin texture. (f) k-space distribution of the trace of the quantum metric tensor ($g_{HH}+g_{VV}$), strongly peaked around the two pseudospin monopoles.}
 \end{figure}

Figure~\ref{fig2} shows the results of the measurements taken at 0~T (no Zeeman splitting $\Delta_z = 0$). The dispersion of the eigenstates extracted from the raw PL data (see Supplementary Note 2) is shown in Fig.~\ref{fig2}(a) (the inset shows the energy difference). At $k=0$, we observe an energy splitting of 16~$\mu$eV between the modes polarised along two perpendicular directions $k_x$ and $k_y$ slightly rotated (see Fig.~\ref{fig2}(b)) with respect to the reference axes $k_H$ and $k_V$ defined by the orientation of the polarizer. 
If this XY splitting were zero ($\alpha=\Delta_z=0$), the dispersion would be composed of two parabola of different masses touching at $k=0$, similar to the quadratic band degeneracies of valley electrons in bilayer graphene. The corresponding effective field (pseudospin)  would stay in the plane showing a dipolar texture in momentum space due to the off-diagonal TE-TM term $\propto k^2 e^{-2i\varphi}$. When the XY splitting is non-zero ($\alpha\neq 0$), as it is in our sample, the cylindrical symmetry is broken. Along the direction $\mathbf{e_y}$ in reciprocal space ($\mathbf{e_y}=\cos\varphi_0\mathbf{e_H}+\sin\varphi_0\mathbf{e_V}$), the lowest energy mode at $\mathbf{k}=\mathbf{0}$ has the highest effective mass, and the splitting between the modes simply increases with $k$. Along the perpendicular direction ($\mathbf{e_x}$), the lowest energy mode the smallest mass and the two parabola cross at a finite $k_0=\sqrt{\alpha/\beta}\approx1.1$~ $\mu$m$^{-1}$, giving rise to two tilted Dirac cones. At this point, the TE-TM splitting perfectly compensates the XY splitting, whereas along $k_y$ such point is absent, since both contributions add up and the splitting grows monotonously. This is clearly visible in Figs.~\ref{fig2}(b,c), showing the extracted polarisation degree of lower energy band in the HV and DA basis (the circular polarisation degree is close to zero at 0~T). These panels exhibit the typical four-lobe structure of the TE and TM modes slightly distorted by the XY splitting. The degeneracy points are marked by a cross, the line passing by them being the $x$ direction. In the vicinity of these points, the in-plane effective field and the pseudospin textures take the shape of 2D monopoles [Fig.~\ref{fig2}(d,e)]. The breaking of the TE-TM rotational symmetry by the XY field induces the splitting of the dipolar TE-TM field into a pair of  2D monopoles with opposite signs but the same winding.  The Berry curvature associated with each of them is a delta function, whereas the metric, even if infinite precisely at these points, has a finite extension in the reciprocal space, decaying as $1/\left(k-k_0\right)^2$. The immediate physical consequence is that the deviation of any experimental measurement of the Berry phase for a loop around these points from the expected value of $\pi$ grows accordingly, when the loop size is reduced. The results of the measurements of this metric are shown in Fig.~\ref{fig2}(f). Each of the effective monopoles can be interpreted as an emergent non-Abelian gauge field acting on photons \cite{Tercas2014}.
 
\begin{figure}
 \includegraphics[width=1.0\linewidth]{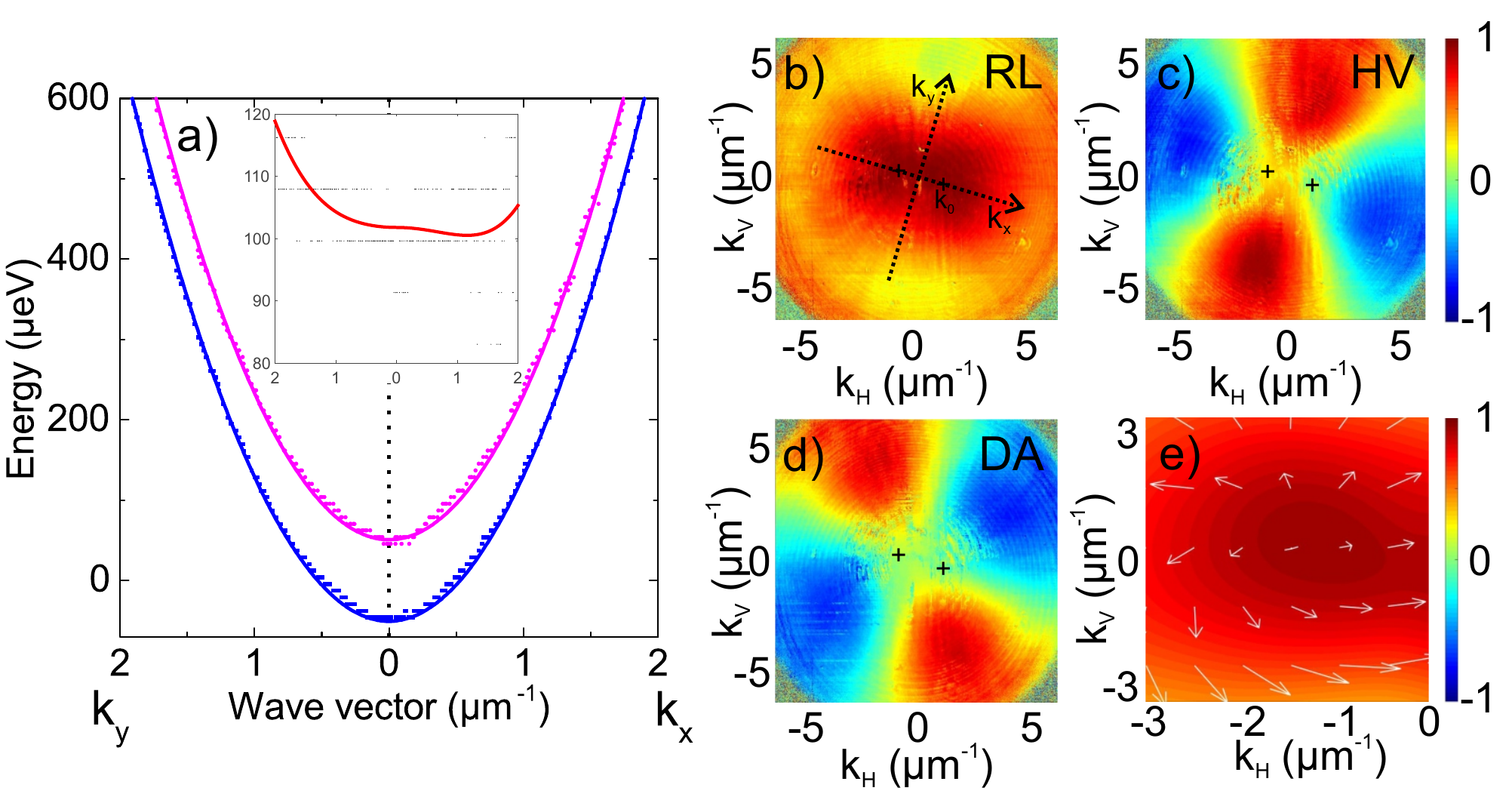}
 \caption{\label{fig3} \textbf{Broken time reversal symmetry: Emergence of half-skyrmion pseudospin textures}
 \small All data presented in this figure are extracted from PL measurements taken at 9T.
  a) Energy dispersion with respect to the two axis of the XY splitting directions (as in Fig. 2(a)). Anticrossing of the branches is observed instead of their crossing. The two polarisation bands are split for all wave vectors (see inset where along the $k_x$ direction the splitting has a non-zero minimum.). (b,c,d) RL, HV, and DA polarisation degrees maps of the lower energy mode. The crosses mark the anti-crossing points. e) Corresponding pseudospin texture distribution in $k$-space, zoomed near one of the crossing points. The in-plane pseudospin $(S_1,S_2)$ is shown by the white arrows. The $S_3$ pseudospin amplitude is shown by colours.}
 \end{figure}

Now we break the time-reversal symmetry by applying a magnetic field of 9~T described by the Zeeman term $\Delta_z$  in the Hamiltonian. Figure~\ref{fig3} shows the results of the measurements. Figure~\ref{fig3}(a) shows the mode dispersions along  $k_x$ and $k_y$, as in Fig.~\ref{fig2}(a) (the inset shows the energy difference). The two modes are now split by 110 $\mu$eV at $k=0$. The crossing along $k_x$ becomes an anticrossing and the two polarisation bands are split everywhere in $\mathbf{k}$-space.
One should notice that despite working at optical frequency, this splitting, appearing thanks to the excitonic fraction of polaritons, is quite significant, being 10 times larger than the linewidth of our ultra-high quality sample.
At the two anti-crossing points, the energy splitting is entirely determined by the Zeeman splitting, giving 100 $\mu$eV.  The measured polarisation degrees are shown in Figs.~\ref{fig3}(b-d). The polarisation at $k=0$ becomes elliptical. The circular polarisation degree decreases along $k_y$ and increases along $k_x$ up to $\pm k_0$, where it becomes close to 1, before decreasing while going towards larger $k$. A zoom on the measured pseudospin texture around $k_0$ is shown in Fig.~\ref{fig3}(e), exhibiting a part of a half-skyrmion, as expected. 

\begin{figure}
 \includegraphics[width=1.0\linewidth]{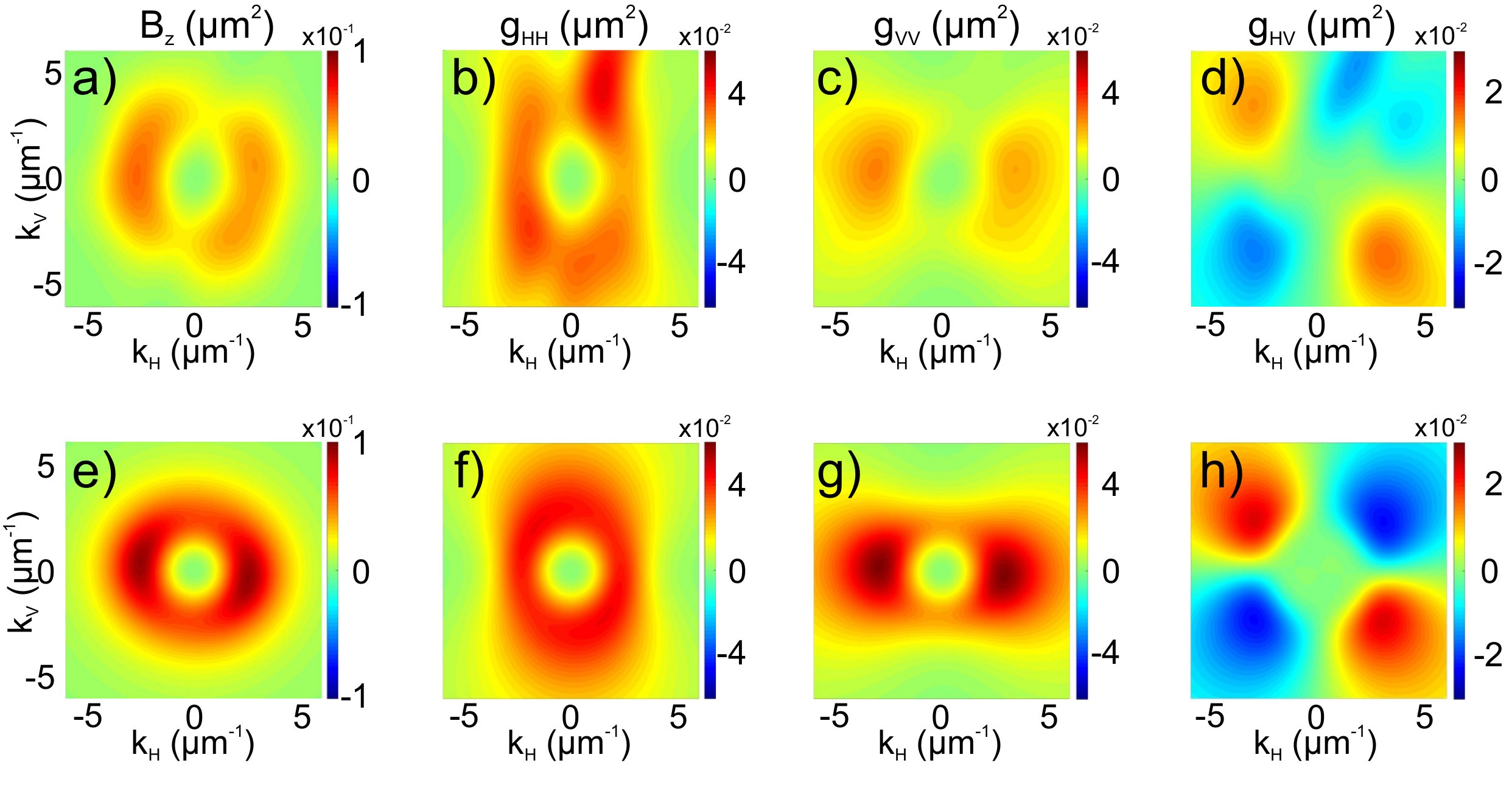}
 \caption{\label{fig4} \textbf{Berry curvature and quantum metric distributions}
 \small  (a-d) - experiment (PL at 9~T). k-space distribution of quantum geometric tensor elements: (a) Berry curvature $B_z$, (b) $g_{HH}$, (c) $g_{VV}$, (d) $g_{HV}$, extracted using Eq.~\eqref{xtract}. (e-h) - theory. The computation is based on the effective Hamiltonian \eqref{Hameff} and analytical expressions given in Supplementary Note 1.}
 \end{figure}

The $k$-space distributions of the Berry curvature and of the three components of the quantum metric tensor, extracted from the experimental data of Fig.~\ref{fig3} using Eq.~\eqref{xtract}, are shown in Figs.~\ref{fig4}(a-d) and compared with analytical predictions (Figs.~\ref{fig4}(e-h), see Supplementary Note 1) based on the parameters $\alpha=8$~$\mu$eV, $\beta=5.92$~$\mu$eV$\mu$m$^2$, $\Delta_z=50.2$~$\mu$eV, $m=2.1\times 10^{-4}m_0$ extracted from the dispersions in Fig.~\ref{fig2}(a),\ref{fig3}(a) (using no fitting parameters, $m_0$ being the free electron mass). Without the XY splitting, the Berry curvature distribution would be doughnut-like (circularly symmetric), whereas for a dominating XY splitting the distribution would be concentrated around the anti-crossing points. Here, we are between these two limiting cases. A similar procedure has been applied to the second polarisation energy branch (Supplementary Figure 4). The measurements confirm that at a given $k$, the two branches are cross-polarised and show opposite Berry curvatures and the same quantum metric elements, as expected for a two-band system. Here we want to stress again that a Zeeman splitting alone would not induce any Berry curvature nor quantum metric, the crucial ingredient for both being the k-dependent TE-TM based  spin-orbit-coupling. \\

The high quality of the experimental images of the Berry curvature and the metric presented in Fig.~\ref{fig4}, and their agreement with theoretical predictions clearly demonstrate the efficiency of the direct measurement of the quantum geometric tensor, which is the central point of the present work. It also highlights the important advantage of photonic and polaritonic systems, where such high-resolution measurements are possible with standard optical techniques. This is one of the cornerstones of analogue physics, which capitalises on the mathematical similarities between systems (such as topological insulators with electrons and photons) to access what would be otherwise inaccessible. The precise direct measurement of both the Berry curvature and the quantum metric is important for possible optovalleytronic applications, allowing to predict the wavepacket trajectories at the device design stage. In our system, the measured Berry curvature would lead to a total anomalous Hall drift of  0.3~$\mu$m in dynamical experiments on wavepacket propagation, which can be observed for 20~$\mu$m propagation. This drift increases while decreasing the Zeeman field, but the price to pay is an increased non-adiabaticity, which we can precisely estimate to be 7\% for this set of parameters thanks to the knowledge of the quantum metric (Supplementary Note 3). Beyond the direct measurement of the band geometry, our results unveil the intrinsic chirality of the TE and TM polarised 2D photonic modes in presence of an effective Zeeman splitting, which is the foundation of topological photonics. Indeed, by using an appropriate lattice, the bulk dispersion transforms into gapped photonic Bloch bands which, due to the Berry curvature, can be associated with non-zero Chern numbers and therefore give rise to gapless chiral photonic edge states \cite{Haldane2008,lu2014topological,Nalitov2015}.
Moreover, our procedure can be applied to measure the Bloch band geometry of photonic lattices \cite{lu2014topological,Jacqmin2014} hosting non trivial geometrical or topological effects \cite{Nalitov2015,klembt2018exciton}. Another layout is offered by the use of polariton quasi-particles (interacting photons) able to demonstrate lasing and interacting quantum fluid behaviour (superfluidity, quantized vortices...). The marriage between polariton physics and topological systems has already brought striking breakthroughs, such as topological lasers \cite{StJean2017}, and offers exciting perspectives for the future.


\begin{methods}
\textbf{Experimental setup}. 
The experimental sample is a high quality $\text{GaAs/Al}_x\text{Ga}_{1-x}\text{As}$ planar microcavity with an excellent quality factor Q $\sim10^5$ and embedding $12$ quantum wells (QW), giving a Rabi splitting of 16 meV. This results in a polariton lifetime longer than 100 ps and a good intensity of the PL of the lower polariton branch also when keeping the density per QW below $1$ polariton per $\mu\text{m}^{2}$, avoiding any nonlinear effect. The measurements were performed at approximately zero detuning \cite{Microcavities}.
The microcavity is cooled to 3 Kelvin in a closed loop helium cryostat equipped with a superconductive magnet able to generate a field onto the sample that spans from -9 to 9 T in a Faraday configuration (external magnetic field perpendicular to the microcavity plane).
The excitation is performed by an off-resonant linearly polarized continuous wave $2~\mu \text{m}$ laser spot, tuned in the first minimum of the stopband oscillations in order to maximize the injection.
The sample excitation and polaritonic PL collection is performed in a reflection scheme, by means of a wide numerical aperture objective (0.86), resulting in a $12~\mu \text{m}^{-1}$ field of view, and the $k$ space is reconstructed on the monochromator slits so that the PL can be energetically resolved.
In order to avoid any loss of information of $k$ space the entire detection line is built in a 2f configuration and the needed polarization filtering is performed in the real space plane. The polarization response of the setup is characterized prior to the experiments.
The raw PL data are collected by an automatized labview routine able to perform a complete tomography in any of the 3 polarization bases (H/V, A/D and R/L), throughout sequential passage of light by a couple of motorized quarter- and half-waveplates and a polarizer.
The energy mapping onto the charge-coupled device (CCD) camera is performed throughout a 550 cm monochromator equipped with a $1800$ lines/mm grating and slits aperture set to $80~\mu \text{m}$. 
For each polarization, a scan of 561 images is acquired, each containing a $I(E,k_y)$ map at a given $k_x$, upon moving a translational stage mounting the final lens by steps of $12~\mu \text{m}$. In this way, a 3D set of PL data $[I(E,k_x,k_y)]$ is collected in any of the 6 polarization states.
The image energy resolution is $\delta E = 0.004~\text{nm} \cdot \text{pixel}^{-1}$. The momentum resolution are $\delta k_y = 0.012~\mu \text{m}^{-1} \cdot \text{pixel}^{-1}$ and $\delta k_x = 0.024~\mu \text{m}^{-1} \cdot \text{frame}^{-1}$, corresponding to the momentum magnification with respect to the CCD pixel size and scan lens movement step, respectively.

\noindent\textbf{Numerical analysis}. We start by fitting the total intensity for each wave vector $(k_H,k_V)$ with a double Gaussian curve, which allows to find the energies of the two eigenstates ($E_{\pm}$). Then, the intensities of the 6 polarisation components are obtained at the energies of the eigenstates by integration within the Gaussian width, and the components of the pseudospin calculated from these intensities. If the modes are almost degenerate in total intensity, with the energy difference falling below the inhomogeneous broadening, they can often still be distinguished by studying the spectra in polarisation components separately. This allows to resolve the branches for small energy differences. The pseudospin maps of the lower and upper eigenstates encoded in the angles $\theta$ and $\phi$ are then smoothed with a low-pass filter eliminating the noise. Finally, the components of the quantum geometric tensor are calculated from the pseudospin with Eqs.~\ref{xtract}. Parallel computing is used to accelerate the treatment of $4.6\times 10^9$ experimental data points. The final resolution of the quantum geometric tensor components is $1024\times 561$. The parameters of the $2\times 2$ Hamiltonian~\eqref{Hameff} $\alpha$, $\beta$, $\Delta_z$ are determined with a relative error of 3\%, 1\%, and 0.5\%, respectively.

\end{methods}

\bibliographystyle{naturemag}
\bibliography{biblio} 

\begin{addendum}
\item We thank David Colas for critical reading of the manuscript. This work was supported by the the ERC project "ElecOpteR" (grant N. 780757). We acknowledge the support of the project "Quantum Fluids of Light"  (ANR-16-CE30-0021), of the ANR Labex Ganex (ANR-11-LABX-0014), and of the ANR program "Investissements d'Avenir" through the IDEX-ISITE initiative 16-IDEX-0001 (CAP 20-25). D.D.S. acknowledges the support of IUF (Institut Universitaire de France).
\item[Author contributions] A.G., L.D., and D.B. designed the setup. 
A.G. realised the experiments with the help of V.A. and M.D.G.. D.S. supervised the experimental part. 
K.W. and L.N.P. fabricated the sample. O.B., D.D.S., and G.M. performed the treatment of the experimental data. O.B. performed analytical calculations. G.M. and O.B. wrote the manuscript with inputs from all authors.
 \item[Competing Interests] The authors declare that they have no
competing financial interests.
 \item[Correspondence] Correspondence
should be addressed to G. Malpuech (E-mail: guillaume.malpuech@uca.fr) and D. Sanvitto (E-mail: daniele.sanvitto@nano.cnr.it).
\end{addendum}

\setcounter{secnumdepth}{0}
\addto\captionsenglish{\renewcommand{\figurename}{Supplementary Figure}}
\addto\captionsenglish{\renewcommand{\refname}{Supplementary References}}

\section{Supplementary Notes}

This supplementary materials is composed of three notes.  The first note presents analytical formula used to compute the quantum metric and Berry curvature and which are taken from \cite{Bleu2018}. In the second note, we present raw Photoluminescence data and the deconvolution procedure used to obtained the experimental data presented in the main manuscript. The third note describes the computation of the anomalous drift and non-adiabatic fraction of an accelerated polariton wave packet as mentioned in the main text. This computation is performed using the Berry curvature and quantum metric found in this manuscript. We also show the experimental images of the Berry curvature and the quantum metric of the upper polarisation mode (fig. 4). As expected, it shows an opposite Berry curvature with respect to the lower polarisation mode presented in the main text.

\clearpage
\section{Supplementary Note 1. Analytical expressions for quantum metric and Berry curvature}
For a given band, the elements of the quantum geometric tensor can be defined as:
\begin{eqnarray}
T^{(n)}_{ij}=\sum_{m\neq n}\frac{\bra{\psi_m}\partial_{k_i}H_k\ket{\psi_n}\bra{\psi_n}\partial_{k_j}H_k\ket{\psi_m}}{(E_m-E_n)^2} \label{THam}
\end{eqnarray}
where $H_k$ is the $k$-dependent Hamiltonian, $\ket{\psi_n}$ and $E_n$ the corresponding eigenstates and eigenenergies ($n$ is the band index).
The quantum metric tensor and the Berry curvature tensor are obtained from real symmetric and the imaginary antisymmetric part of \eqref{THam}:
\begin{eqnarray}
g^{(n)}_{ij}=\Re\left[T^{(n)}_{ij}\right],~ \Omega^{(n)}_{ij}=-2\Im\left[T^{(n)}_{ij}\right]=i\left(T^{(n)}_{ij}-T^{(n)}_{ji}\right)
\end{eqnarray}
For a two-band system, these definitions imply that the metric elements of each band are equal $g^{(1)}_{ij}=g^{(2)}_{ij}$ whereas the Berry curvature ones are opposite.
In three dimensional space, the Berry curvature can be represented as a pseudo-vector $\mathbf{B}^{(n)}=\left(\Omega^{(n)}_{23},\Omega^{(n)}_{31},\Omega^{(n)}_{12}\right)^T$. Since the system under study is a planar structure, the only non-zero component of $\mathbf{B}^{(n)}$ is out of the plane, along the z-direction.

Starting with the effective Hamiltonian presented in the main text, one can obtain the analytical expressions of the quantum geometric tensor elements:
\scriptsize
\begin{eqnarray}
g_{HH}&=&\frac{\beta^2\left( \alpha^2 \left[\left( k_V^2-k_H^2\right)\cos2\varphi_0 +2 k_H k_V \sin2 \varphi_0+k^2\right]-4\alpha\beta k^2\left(k_V^2 \cos\varphi_0+k_H k_V \sin\varphi_0\right)  + 2\beta^2 k_V^2 k^4+\Delta_z^2 k^2\right)}{2\left(\alpha^2+2\alpha\beta \left((k_H^2-k_V^2)\cos\varphi_0-2k_H k_V \sin\varphi_0 \right) +k^4\beta^2 +\Delta_z^2 \right)^2}\\
g_{VV}&=&\frac{\beta^2\left( \alpha^2 \left[ \left( k_H^2-k_V^2\right)\cos2\varphi_0 +2 k_H k_V \sin2 \varphi_0+k^2\right]+4\alpha\beta k^2\left( k_H^2 \cos\varphi_0 -k_H k_V \sin\varphi_0 \right)  + 2\beta^2 k_H^2 k^4+\Delta_z^2 k^2\right)}{2\left(\alpha^2+2\alpha\beta \left((k_H^2-k_V^2)\cos\varphi_0-2k_H k_V \sin\varphi_0 \right) +k^4\beta^2 +\Delta_z^2 \right)^2}
\\
g_{HV}&=&\frac{\beta^2\left( \alpha^2 \left[ \left( k_H^2-k_V^2\right)\sin2\varphi_0 +2 k_H k_V \cos2 \varphi_0\right]+2\alpha\beta k^4\sin\varphi_0- 2\beta^2 k_Hk_V k^4\right)}{2\left(\alpha^2+2\alpha\beta \left((k_H^2-k_V^2)\cos\varphi_0-2k_H k_V \sin\varphi_0 \right) +k^4\beta^2 +\Delta_z^2 \right)^2}
\\
B_z^{\pm}&=&\frac{\pm2\beta^2k^2\Delta_z}{\left( \Delta_z^2+\left(\left(k_H^2-k_V^2 \right)\beta+\alpha \cos \varphi_0\right)^2+\left(-2k_H k_V\beta+\alpha \sin\varphi_0  \right)^2\right) ^{3/2}} 
\label{xyQGT}
\end{eqnarray}
\normalsize
In the figure 4 of the main text, we use the above expressions for a direct comparison with the experimental extraction, 
the value of $\alpha$, $\beta$ and $\Delta_z$ being extracted from the experimental data. The slight angle $\varphi_0$ between the constant in-plane field and the experimental HV axis complexify a bit the equation with respect to the ones presented in \cite{Bleu2018} which are recovered by taking $\varphi_0=0$.

\clearpage

\section{Supplementary Note 2. Raw Photoluminescence data and its treatment}
\begin{figure}[h!]
 \includegraphics[width=1.0\linewidth]{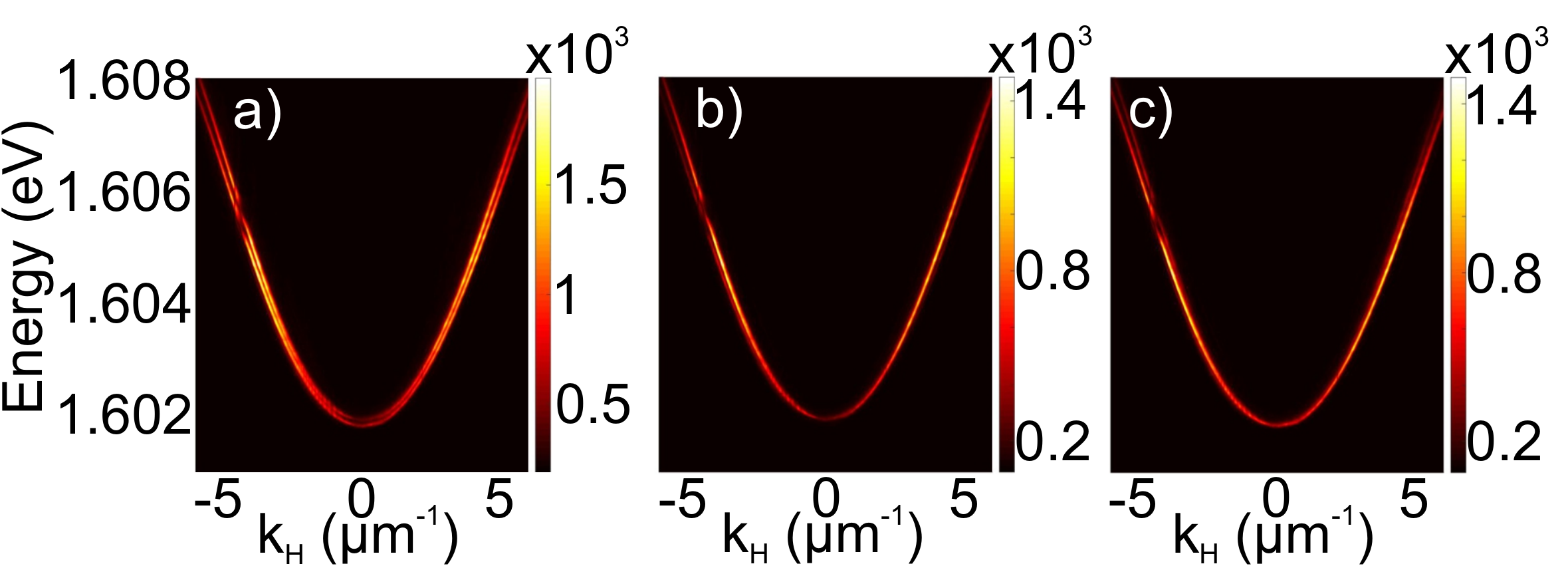}
 \caption{\label{figS1} Measured photoluminescence intensity as a function of energy and wave vector at 9 Tesla: a) total, b) left-circular, c) right-circular.
 }
 \end{figure}

The Supplementary Figure~\ref{figS1} shows the raw PL spectra versus $k_H$ measured at 9~T: (a) - total intensity, (b) - left-circular, (c) - right-circular. One can see that the lines are well resolved, with the splitting exceeding their broadening (of all origins) everywhere. They show a substantial circular polarisation degree near $k=0$. 

\begin{figure}[h!]
 \includegraphics[width=1.0\linewidth]{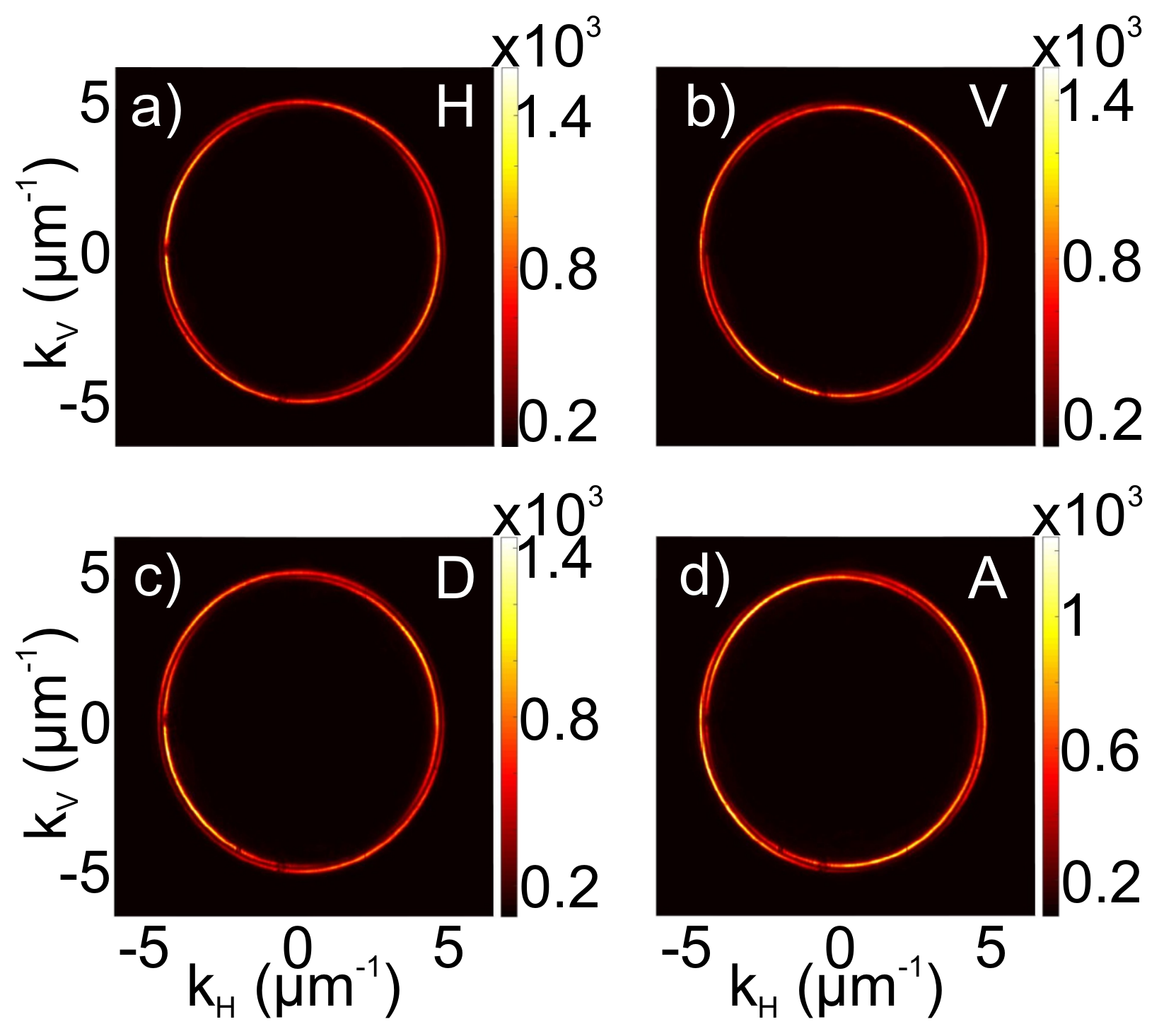}
 \caption{\label{figS2} Measured photoluminescence intensity as a function of in-plane wave vector, 4.5 meV above the ground state energy and under 9 Tesla: a) horizontal, b) vertical, c) diagonal, d) anti-diagonal.
 }
 \end{figure}

The Supplementary Figure~\ref{figS2} shows the photoluminescence intensity measured again at 9~T in the four linear polarisation projections: horizontal (a),  vertical (b), diagonal (c), and anti-diagonal (d),  versus the wave vector $k$ at a given energy (4.5 meV above $k=0$), where the mode structure is governed by the TE-TM splitting. One clearly sees the presence of two interleaved circles. For each direction in the reciprocal space, the polarisations of the inner and outer circles are orthogonal. The polarisation patterns,  presented in the main text (Fig.~2(b,c), Fig.~3(b,c,d)) are extracted from these types of measurements. 

One should note, however, that the energy difference between modes is quite smaller at 0T, where the modes are even crossing at two points in reciprocal space. The Supplementary Figure~3(a) shows the total intensity taken at 0T along $k_H$ respectively. One can see that the mode splitting close to $k=0$ is not straightforwardly resolved since the splittings are comparable with the linewidth. This is imposing the use of a standard deconvolution procedure to extract the two modes energy and polarisations for each wave vector. We ultimately used this procedure to treat all the sets of data. 

The measured k-space polarisation patterns of the main text, even if very clear, show some short scale fluctuations due to the experimental setup. These fluctuations would induce errors in the computation of the quantum geometric tensor elements, which require taking derivatives. We therefore use a low-pass spectral filtering procedure. The procedure is based on the 2D Fourier transform of the k-space polarisation, removal of the fluctuations, and inverse Fourier transform. This allows to get a pseudo spin texture free cleaned of experimental noise (Fig.~2(d,e), Fig.~3(e)) and then to compute the different elements of the quantum geometric tensor shown in Fig.~2(f) and Fig.~4(a-d).

\clearpage

\section{Supplementary Note 3. Semi-classical computation of anomalous Hall drift and non-adiabatic fraction}

In this supplementary note, we provide an illustration of the use of the Berry curvature and quantum metric measured in the main text by showing a computation of the anomalous Hall drift of an accelerated wave packet. First, we provide a simple analytical estimate based on the monopolar nature of the gauge field arising in the reciprocal space because of the interplay of the constant XY field with the $k$-dependent TE-TM effective field. Then, we carry out a semi-classical calculation of the anomalous Hall drift, which allows us to obtain a more precise value and to check that the evolution is sufficiently adiabatic, meaning that the approximation remains correct.

A monopolar in-plane effective field combined with a Zeeman splitting is equivalent to the well-studied configuration of the 2D Dirac Hamiltonian, which describes not only relativistic electrons and positrons in 2D, but also many analog systems, in particular, the wide class of 2D materials (graphene, transitional metal dichalcogenides). The maximal anomalous Hall drift due to the Berry curvature in such systems can be simply estimated as the Compton electron wavelength, which is the combination of parameters of the Dirac equation with the dimensions of length:
\begin{equation}
\lambda_{Compt}=\frac{\hbar}{mc}
\end{equation}
The parameters $m$ and $c$ in our case should be chosen to describe the dispersion at the anticrossing point $k_0=\sqrt{\alpha/\beta}$: the slope of the dispersion $c$ is given by $\hbar c=\sqrt{\beta\alpha/2}$, and the mass corresponds to the splitting of the branches $mc^2=\Delta_z$, which gives the estimation for the maximal anomalous Hall drift:
\begin{equation}
\Delta Y_{AHE}=\frac{\sqrt{\beta\alpha}}{\Delta_z\sqrt{2}}
\end{equation}
This expression can be compared with the anomalous Hall drift without the in-plane field found in \cite{Bleu2018effective}:
\begin{equation}
\Delta Y_{0}=\frac{\sqrt{\beta}\Gamma^2(3/4)}{\sqrt{\Delta_z}\sqrt{\pi}}
\end{equation}
An important difference is that the dependence on the Zeeman splitting changes: it is stronger in presence of the in-plane field $\alpha$, because the Berry curvature becomes completely concentrated in the anticrossing points, and not distributed along a ring in the reciprocal space.

With the parameters extracted from the experimental dispersion in the main text $\alpha=8$~$\mu$eV, $\beta=5.92$~$\mu$eV$\mu$m$^2$, $\Delta_z=50.2$~$\mu$eV, we obtain $\Delta Y_{AHE}\approx 0.3$~$\mu$m.  Decreasing the Zeeman splitting by a factor 2 allows obtaining $\Delta Y_{AHE}=0.6$~$\mu$m.

The expressions for the anomalous Hall drift  given above are actually based on the semi-classical equations for the propagation of a wave packet in presence of the Berry curvature. These equations are based on the approximation of adiabatic evolution. They can include corrections accounting for the non-adiabaticity. If the evolution occurs along geodesic trajectories in the parameter space, the semi-classical equations including the dominant correction read \cite{Bleu2018effective}:
\begin{eqnarray}
\label{semi}
  \hbar \frac{{\partial {\mathbf{k}}}}{{\partial t}} &=& {\mathbf{F}} \\ 
  \hbar \frac{{\partial {\mathbf{r}}}}{{\partial t}} &=& \frac{{\partial {\mathbf{\tilde E}}}}{{\partial k}} + \hbar \frac{{\partial {\mathbf{k}}}}{{\partial t}} \times {\mathbf{B}}\left( {1 + 2\frac{{{g_{xx}}}}{{{\Omega ^2}}}{{\left( {\frac{{\partial {\mathbf{k}}}}{{\partial t}}} \right)}^2}} \right)
\end{eqnarray}
Here, we assume that the acceleration occurs along the $X$ axis (the one with the anticrossing point); $\mathbf{B}(\mathbf{k})$ is the Berry curvature, $F$ is the accelerating force (for example, from an energy gradient), $g_{xx}(\mathbf{k})$ is the quantum metric component corresponding to the acceleration direction, $\tilde{E}(\mathbf{k})=(1-g_{xx}(\mathbf{k})/\Omega^2(\mathbf{k}))E_0(\mathbf{k})+g_{xx}(\mathbf{k})/\Omega^2(\mathbf{k}) E_1(\mathbf{k})$ is the corrected energy for the calculation of the group velocity, and $\Omega(\mathbf{k})$ is the total splitting between the branches.

The non-adiabatic fraction
\begin{equation}
f_{NA}=\frac{g_{xx}}{\Omega^2}
\end{equation}
allows to find the fraction of the system present in the excited state at any moment of time. For the adiabatic approximation to hold (even with corrections), this fraction should be sufficiently low during the whole evolution. That's why it is important to monitor the maximal value of $f_{NA}$.

The results of the calculations of the wavepacket trajectory based on Eq.~\eqref{semi} are shown in Fig.~\ref{figSsemi}. The black solid curve corresponds to the trajectory of a wavepacket in a cavity with the parameters of the one given in the main text and above, with a potential gradient of $1$~meV/$20$~$\mu$m, taking into account the non-adiabatic correction. The maximal value of the non-adiabatic fraction is 7\% in this case, and the maximal displacement of about 0.3~$\mu$m corresponds well to the estimation above. The red dashed curve is calculated without the non-adiabatic correction. It is shown here as a reference. 

It is possible to increase the anomalous Hall drift by reducing the Zeeman splitting. However, it also requires decreasing the potential gradient and thus the accelerating force, in order to keep the non-adiabatic fraction constant.
The solid green curve exhibits a twice higher anomalous Hall drift of about 0.6~$\mu$m thanks to a reduced Zeeman splitting of $20$~$\mu$eV, compensated by a lower gradient of $1$~meV/$70$~$\mu$m, giving a maximal non-adiabatic fraction of 8\%.

\clearpage 
\section{Supplementary figures}
\clearpage 

\clearpage 

\begin{figure}[h!]
 \includegraphics[width=1.0\linewidth]{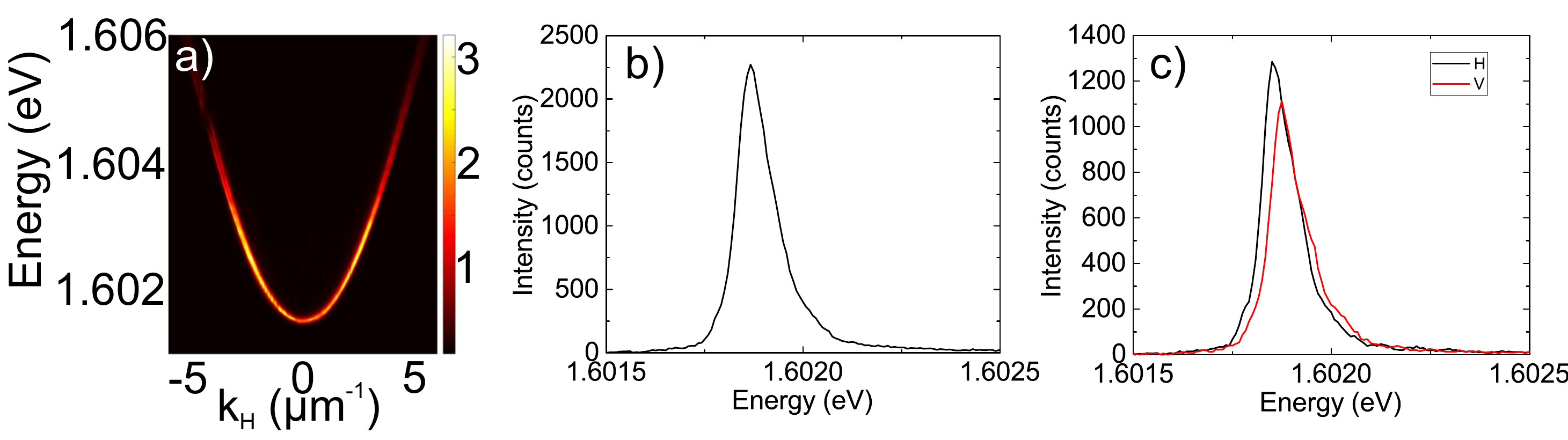}
 \caption{\label{figS3} Measured photoluminescence intensity at 0 Tesla: a) total intensity as a function of wave vector and energy, b) total intensity as a function of energy only, for wave vector $k_H=1.46$~$\mu$m$^{-1}$; c) intensities in H and V polarisations for the same wave vector.
 }
 \end{figure}
 
 \clearpage

\begin{figure}[h!]
 \includegraphics[width=1.0\linewidth]{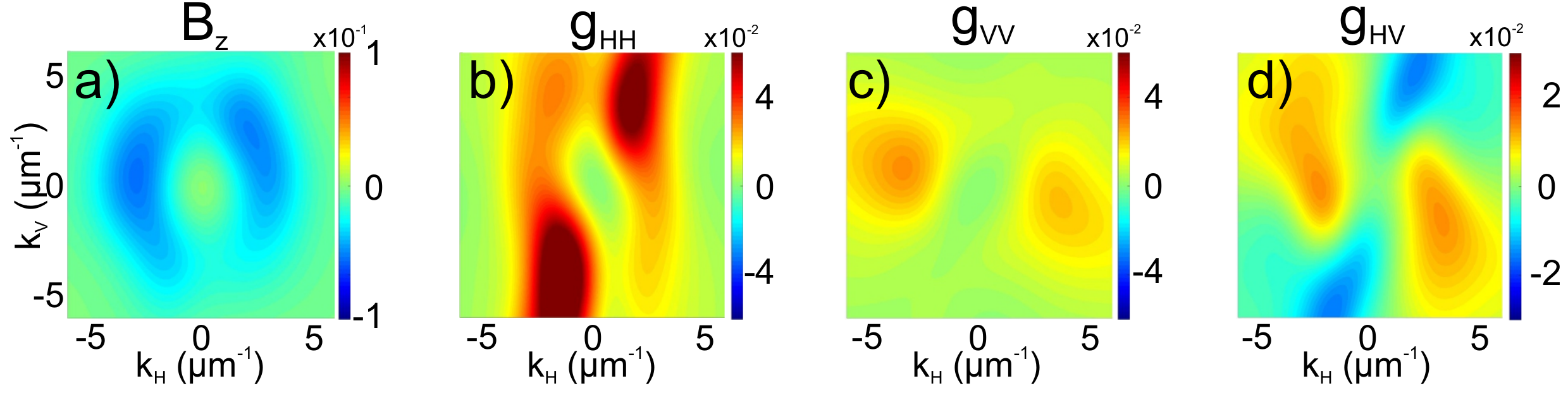}
 \caption{\label{figS5} Berry curvature and quantum metric for the upper branch, measured at 9~T. k-space distribution of quantum geometric tensor elements: (a) Berry curvature $B_z$, (b) $g_{HH}$, (c) $g_{VV}$, (d)$g_{HV}$, extracted using Eq.~(4) of the main text. 
 }
 \end{figure}
 
\clearpage

\begin{figure}[h!]
 \includegraphics[width=1.0\linewidth]{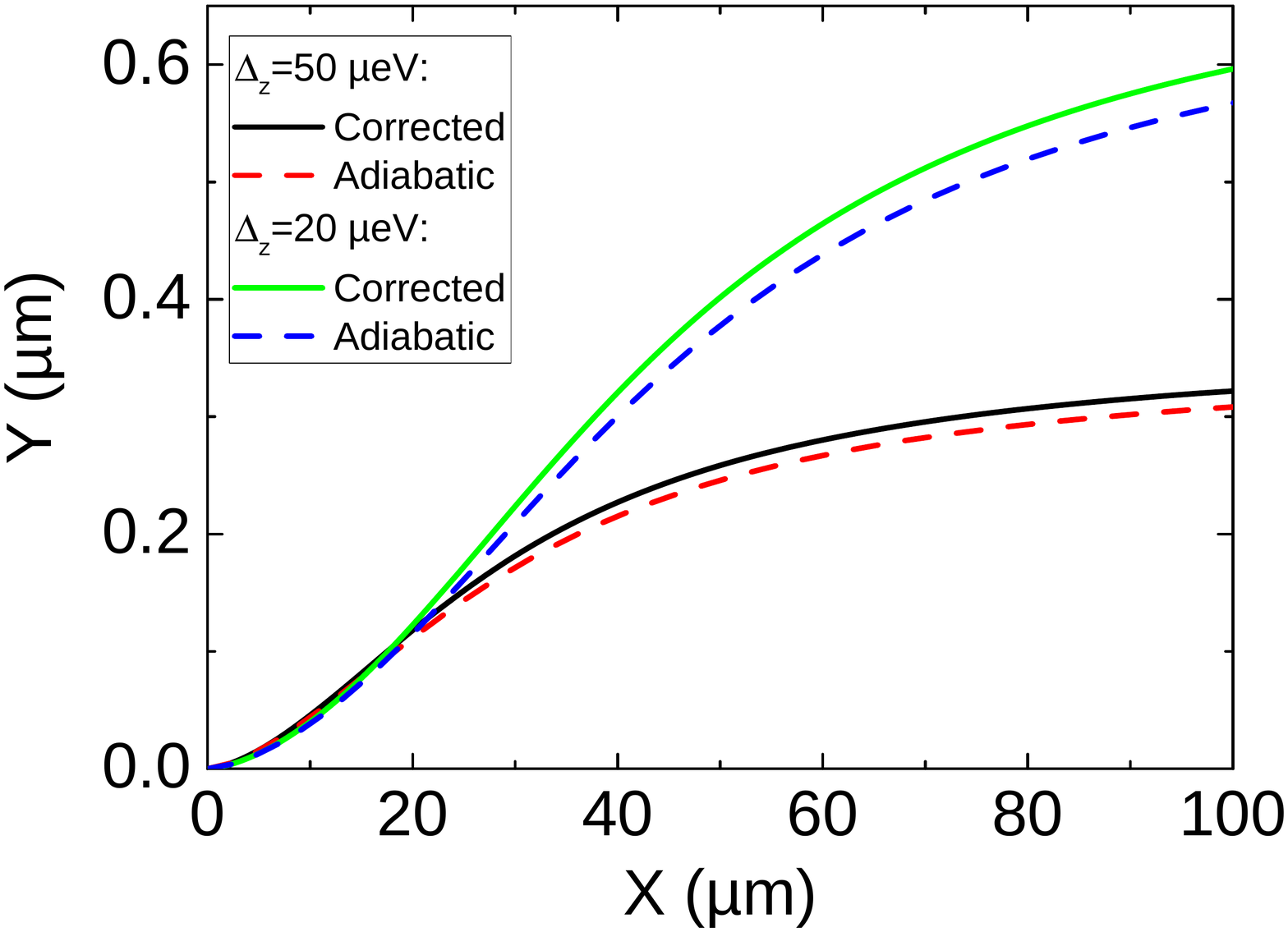}
 \caption{\label{figSsemi} Semi-classical wavepacket trajectories: corrected (solid) and uncorrected (dashed) for two different values of Zeeman splitting ($\Delta_z=50$ and $20$~$\mu$eV) and two different gradients ($1$~meV/$20$~$\mu$m and $1$~meV/$70$~$\mu$m), giving non-adiabatic fractions of 7\% and 8\%, respectively. 
 }
 \end{figure}

\end{document}